%% file: MAIN.tex
\documentclass[sigconf]{acmart}
\usepackage{enumitem}
\usepackage[table]{xcolor}   % enables row/column colors
\usepackage{booktabs}        % nice horizontal rules
\usepackage{tabularx}        % for flexible X column
\usepackage{tablefootnote}
\usepackage[most]{tcolorbox}
\usepackage{fontawesome}
\usepackage{balance}

\tcbset{
    mybox/.style={
        colframe=green!75!black, %
        colback=green!5!white,   %
        coltitle=white,          %
        fonttitle=\bfseries,     %
        breakable,
        arc=0mm,width=\columnwidth,
        top=1mm,left=1mm,  right=1mm, bottom=1mm,
        boxrule=1pt,
    }
}

\definecolor{BrandPink}{HTML}{d0f7ea} % or RGB model: {216,27,96}ck

\AtBeginDocument{%
  }

\usepackage{pifont}      % gives \ding
\newcommand{\TAquote}[3]{%
  \begin{list}{}{%
      \setlength\leftmargin{0pt}%
      \setlength\rightmargin{0pt}}%
    \item\itshape
      \ding{#1}#2%
  \end{list}%
}
\newcommand{\sectopic}[1]{\vspace{0.2em}\par\noindent{\textit{\bfseries #1}}}

\newcommand{\countryName}{{$Australia$}}
\newcommand{\uniName}{{$Monash$}}

\copyrightyear{2026}
\acmYear{2026}
\setcopyright{cc}
\setcctype{by-nc-nd}
\acmConference[ICSE-SEET '26]{2026 IEEE/ACM 48th International Conference on Software Engineering}{April 12--18, 2026}{Rio de Janeiro, Brazil}
\acmBooktitle{2026 IEEE/ACM 48th International Conference on Software Engineering (ICSE-SEET '26), April 12--18, 2026, Rio de Janeiro, Brazil}
\acmPrice{}
\acmDOI{10.1145/3786580.3786977}
\acmISBN{979-8-4007-2423-7/2026/04}

\begin{document}

% \title{They Call Her ‘Miss’ and Him ‘Professor’: A Lived Experiences Report on Gendered Challenges in Computing Education}
\title{They Call Her ‘Miss’ and Him ‘Professor’: Lived Experiences of Women Teaching Support Staff in IT/SE Education}

\author{Vasudha Malhotra}
\email{vasu.malhotra@monash.edu}
\affiliation{%
  \institution{Monash University}
  \city{Melbourne}
  \state{Victoria}
  \country{Australia}
}

\author{Rhea Dsilva}
\email{rhea.dsilva@monash.edu}
\affiliation{%
  \institution{Monash University}
  \city{Melbourne}
  \state{Victoria}
  \country{Australia}
}

\author{Rashina Hoda}
\email{rashina.hoda@monash.edu}
\affiliation{%
  \institution{Monash University}
  \city{Melbourne}
  \state{Victoria}
  \country{Australia}
}

\renewcommand{\shortauthors}{Malhotra et al.}

\begin{abstract}
 Despite their critical role in shaping student learning in computing education, the contributions of women teaching-support staff (TSS) often go unrecognised and undervalued. In this experience report, we synthesise lived experiences of 15 women TSS in IT/SE higher education to illuminate how authority is earned, resisted, and maintained in everyday teaching. Participants shared both their positive and negative lived experiences associated with \textit{finding} and \textit{losing voice} with teaching team colleagues on the one hand, and \textit{rewarding connections} and \textit{gendered friction} with students on the other.  We map these dynamics onto an intersectional ``wheel of privilege and power'' tailored to TSS roles. The farther a TSS profile sits from the wheel's center (e.g., non-native English, non-white, younger-seeming, non-permanent, early-career), the more relational, emotional, and disciplinary labour is needed to reach parity. We provide actionable insights and recommendations for creating more inclusive education environments in technology dominant fields that are particularly timely as universities worldwide grapple with post-pandemic teaching models and seek to build more inclusive and resilient academic communities.
\end{abstract}





\maketitle

\input{Sections/introduction}

\input{Sections/study}

\input{Sections/discussion}

\input{Sections/recommendations}
\input{Sections/background}
\input{Sections/threats}

\input{Sections/conclusion}

%\newpage
\balance
\bibliographystyle{ACM-Reference-Format}
\bibliography{sample-base}


\end{document}

%% file: Sections/introduction.tex
\section{Introduction}~\label{sec:Introduction}

\TAquote{125}{
Some students call me ``Miss'' like they would a kindergarten teacher, even though I hold a PhD. [...] it is quite disturbing to be called ``Miss'' while the male colleague, despite lacking a PhD, is referred to as ``Professor''.
\ding{126}quote from  Participant\#13 in this study}.

The persistent gender imbalance in software engineering (SE), computer science (CS), and information technology (IT) education -- collectively referred to as SE/IT education in this paper -- is one of the chronic challenges facing tertiary education and the digital workforce. Despite multi-decade initiatives, from outreach programmes and coding camps to institutional equity charters, women continue to be markedly under-represented, constituting roughly 28\% of the global STEM workforce and an even smaller share in socio-technical areas such as software development, DevOps, and AI engineering~\cite{murciano2022missing,george2024bridging}. This under-representation begins early, through classroom interactions and assessment practices that shape whether women persist in, or exit from, computing pathways. 

Of the various roles studied in this context, Teaching Support Staff (TSS), often employed as casual or fixed-term teaching assistants (TAs) and technical lab assistants\footnote{We use the term \textit{teaching support staff (TSS)} to acknowledge and respect how the participants preferred to refer to their roles instead of using the term TAs or tutors in our context. The roles and responsibilities of TSS might vary in other contexts.} are a critical yet often overlooked stakeholder in this educational ecosystem~\cite{perlmutter2023field,barkhuff2025exploring}. 
In SE and IT courses, TSS serve as the primary point of contact for students, delivering tutorials, running labs, debugging code, mediating online forums, grading assignments, and often serving as role models for aspiring technologists. For women TSS in these fields, this role presents unique challenges and opportunities~\cite{tari2018someone,khazan2019examining,evans2024gender}. They simultaneously navigate the demands of technical expertise and pedagogical effectiveness, and gendered expectations in male-dominated learning spaces. Their experiences offer a crucial window into understanding how gender bias operates within IT/SE education and how it might be addressed.

While considerable research has examined the experiences of women students in STEM fields and women faculty members~\cite{wang2015gender, google2016diversity, sax2017anatomy}, there exists a significant gap in understanding the lived experiences of women in teaching support roles, particularly in SE contexts~\cite{mirza2019undergraduate}. This oversight is particularly concerning given that TSS positions often serve as entry points into academic careers, influence thousands of students’ first experiences in programming, and can significantly influence decisions about pursuing advanced degrees or academic positions. Women TSS face a unique confluence of challenges which remain understudied~\cite{tari2019understanding}. While the struggles of TSS in computing courses have been covered to some extent, the focus remains on the interactions with students and their tasks in general~\cite{riese2021challenges}. In addition to this wider context, we were motivated to study the experiences of women TSS as it was identified as a pain point in the case studied, detailed in Section~\ref{sec:study}.

In this paper, we present an experience report: a synthesis of first-hand accounts from women TSS about day-to-day teaching in SE/IT higher education. Our approach is exploratory, descriptive, and practice-oriented. We conducted 15 in-depth, semi-structured interviews with women TSS. We foreground their lived experiences —what helped, what hurt, and what patterns recur—using the speakers’ own words (quotations) where possible. We describe and interpret those lived experiences through an equity diversity and inclusion (EDI) lens, asking: \textbf{\emph{What are the lived experiences of women teaching support staff in IT/software engineering higher education, viewed through an EDI lens?}} 

To organise the narratives, we used socio-technical grounded theory for data analysis (STGT4DA), including hashtag coding and memoing \cite{hoda2024qualitative}  to surface key concepts and categories where social dynamics and technical contexts intertwine. Our contributions are: 

 \begin{itemize}[leftmargin=*]
     \item We provide one of the first detailed account of women TSS experiences in SE/IT higher education, those who directly shape the learning experiences of thousands of SE/IT students, revealing major issues that characterise their professional lives and the compounding effects of intersectional identities.
     
     \item Our analysis extends understanding of how gender operates in SE/IT educational settings and reveals intersectionality with a wheel of power and privilege that explains \emph{when and why} the women TSS' authority is discounted in SE/IT education settings.

     \item We provide recommendations that can be applied by institutional leaders, department heads/chairs, course coordinators, individual educators, students, and TSS themselves, both women and men. Our findings have direct relevance for universities seeking to improve gender equity in STEM fields.
 \end{itemize}

\sectopic{Paper Organisation}: Section~\ref{sec:study} covers the experience report context, motivation, and  methodology details. Section~\ref{sec:Findings} presents the main findings from the women TSS lived experiences and Section~\ref{sec:Recommendations} provides recommendations based on our findings. Section~\ref{sec:Background} covers the related work. Section~\ref{sec:threats} present the limitations and threats to validity and Section~\ref{sec:Conclusion} concludes the experience report.

%% file: Sections/study.tex
\section{Experience Report Overview}\label{sec:study}

Below, we share details of the experience report context, motivation, and details of the data collection and data analysis leading to the key findings and actionable insights and recommendations.

\subsection{Context and Motivation}
This experience report is situated in the Faculty of IT (FIT) (over 300 academic staff) at Monash University, Australia. FIT delivers undergraduate (UG) and postgraduate (PG) programs in software engineering, computer science, information systems, artificial intelligence, cybersecurity and related areas -- collectively referred to as technology education in this paper. Teaching includes online or recorded lectures, interactive workshops with lecturing components and hands-on activities and applied classes, e.g., labs or tutorials. Workshops typically host 120-240 students (with two or more team members teaching collaboratively) and the applied sessions typically host 30–60 students (with one or more team members  teaching collaboratively).

The classrooms are generally staffed with teaching teams, including, professors, lecturers and teaching support staff (TSS) such as casual/fixed-term teaching assistants (TAs), lab demonstrators, and administrative/assessment TAs, who provide on-the-spot help, moderate forums, and grade in-class activities as well as rubric-based assignments with calibration meetings. Cohorts are somewhat gender-skewed towards men (typically 25\% women) and include a sizeable international intake, with many students for whom English is an additional language. Within the teaching teams, gender representation is mixed ($\approx$30\% women).

We were motivated to study this topic due to the following reasons. In the end-of-semester debriefs, calibration/remark meetings, ad-hoc reports, and everyday corridor conversations across the Faculty, some patterns were noticed. Reports of disproportionate push back on marks when the assessor was a woman, slower compliance with classroom management requests from women TSS, and more contentious academic-integrity standoffs with women TSS were common. The Dean of the Faculty (a woman) suggested an exploration of these issues. The suggestion was taken up by the Associate Dean EDI (a woman of colour) and the last author. When brought up to the EDI committee, members (men and women) shared similar experiences, either personally encountered—or repeatedly handled reports of—similar issues, and not just at \uniName, but also other institutes and different countries that they had worked in, recognising a wider aspect to this issue. It was clear that women TSS were the most vulnerable. This led to the AD-EDI launching a study into this issue with the intention to share findings with the wider research community and to develop actionable insights and recommendations to improve the state of practice.

\subsection{Women TSS Recruitment}
After gaining ethics approval (Monash ethics application number 44536), we conducted in-depth interviews with 15 women TSS across courses, spanning programming, software project management, usability, and software testing. Participants (referred to as P1--P15) varied in career stage, prior industry experience, and language background. 
Recruitment was purposive with some snowballing, and participation was voluntary (see Table~\ref{table:experience_level_specialisation} for details). We began with purposive sampling, targeting women TSS in SE/IT courses in the Faculty. Participants were recruited through the EDI network and by publicising the study via the Faculty digital newsletter and physical flyers. Two participants were recruited through participant referrals. We intentionally sought diversity in career stages (from first-time TSS to experienced ones), and cultural and linguistic backgrounds. To respect the time spent by women TSS on sharing their valuable experiences with us, we offered them gift vouchers after the interviews.

\input{Tables/tab_participants}

\subsection{Data Collection}
 We developed a pre-interview questionnaire \cite{hoda2024qualitative} to gather demographic information, teaching experience, and institutional context. This allowed us to better understand participants' backgrounds and tailor the interview questions to their specific contexts. We developed a flexible interview guide exploring four main areas (listed below), without restricting participants' narratives. We also asked situational and open-ended probes to encourage participants to elaborate freely on experiences they perceived as significant. Care was taken to avoid prompting respondents towards specific issues or assumptions, allowing their lived experiences to emerge naturally and authentically.

\begin{itemize}[leftmargin=*]
\item \textbf{Background and teaching/industry experience:} Participants' pathways into SE/IT teaching, previous experiences, motivations for becoming TSS, and career trajectory.
\item \textbf{Classroom experiences:} Interactions with students, teaching challenges, establishing authority and credibility, and navigating student expectations.
\item \textbf{Workplace experiences:} Relationships with colleagues, team dynamics, recognition of their work, and distribution of teaching-related tasks.
\item \textbf{Institutional experiences:} Availability and effectiveness of support systems, institutional policies, professional development opportunities, and perceived career progression.
\end{itemize}

Interviews lasted 45-90 minutes and were conducted via Zoom to accommodate participants' schedules. All interviews were audio-recorded with consent. The interviews began by asking participants to describe their pathways into technical teaching and their current roles. This opening allowed participants to establish context before exploring more sensitive topics around gender dynamics. We then used open-ended questions to explore their experiences, following up on specific incidents or observations they shared. The list of pre-interview questionnaire and interview questions can be found in the supplementary materials\footnote{https://tinyurl.com/muy5cvxt}.

\subsection{Data Analysis}~\label{subsec:ResearchMethod}
All interview recordings were transcribed using automated transcription services of Zoom, and manually verified and corrected for accuracy. We removed names and other potentially identifying details at this stage and assigned participant IDs (P1–P15).
All transcripts were then analysed using Socio-Technical Grounded Theory (STGT)~\cite{hoda2021socio,hoda2024qualitative}. We used STGT for analysis, as our analysis aligns closely with its socio-technical research framework (as done by others~\cite{gama2025socio,gunatilake2025role, smithers2025decasualisation}), and STGT provides flexibility to perform analysis at different levels of application (basic vs advanced), including, our basic application of STGT for data analysis to capture TSS lived experiences and present practice-oriented insights.

\subsubsection{Data Analysis Procedure}
STGT involves two stages: a basic stage, dedicated to systematic data collection and initial analysis, and an advanced stage, intended for theory development and refinement~\cite{hoda2024qualitative}. STGT offers researchers flexibility in selecting whether to pursue only the basic stage or progress to the advanced stage based on their research objectives and analytic insights. We employed the basic stage of STGT, as our primary aim was to richly document and explore the classroom, workplace, and institutional experiences of women TSS through an EDI lens and derive actionable recommendations, rather than explicitly developing a new substantive theory.
Below, we describe our analysis procedures.

\begin{figure}[t]
\centering
\includegraphics[width=0.5\textwidth]{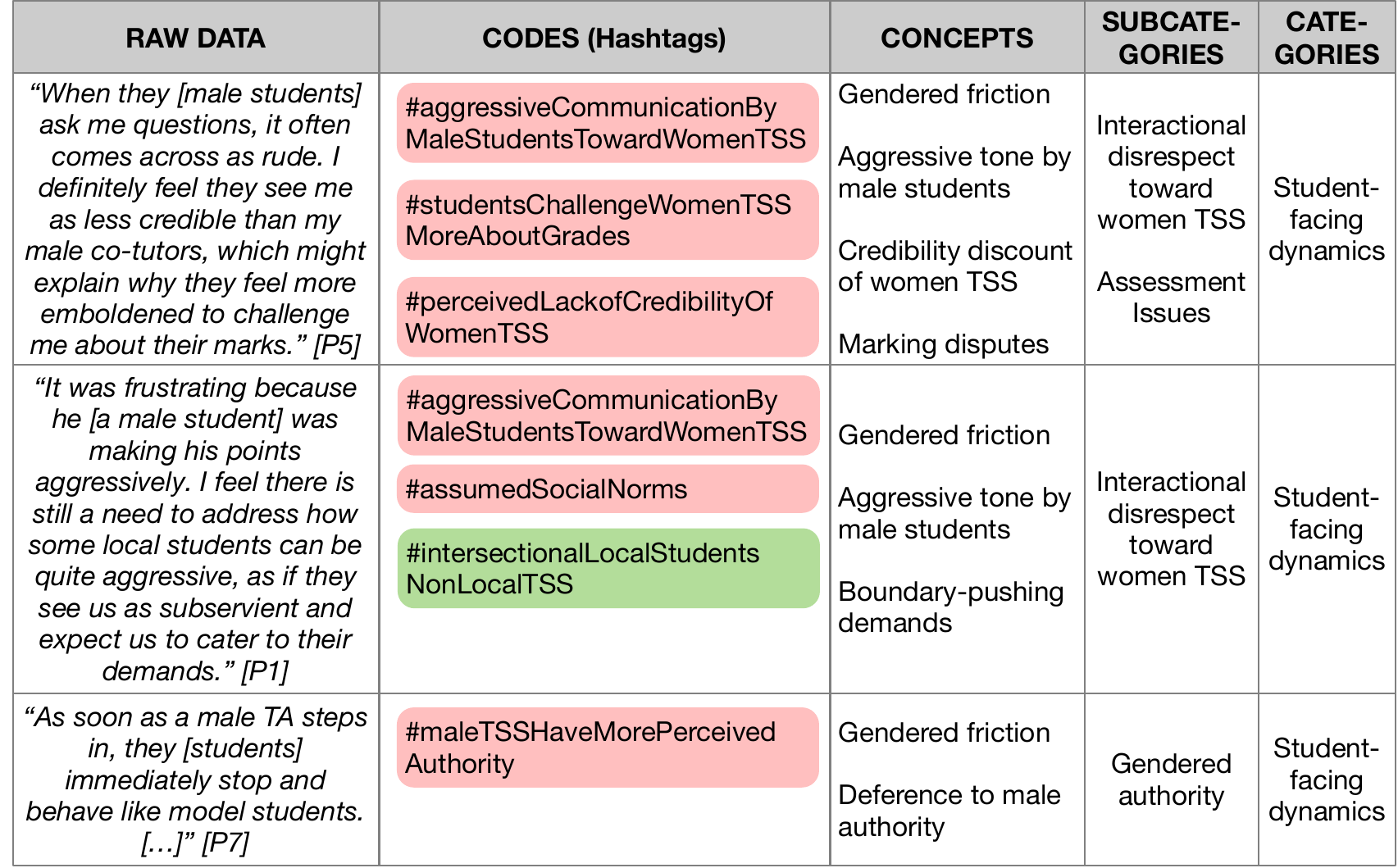}
\vspace*{-1.5em}
\caption{Examples of STGT for Data Analysis}
\label{fig:ExampleCoding}
\vspace*{-1.5em}
\end{figure}

\begin{enumerate}[leftmargin=*,itemsep=2pt]
\item \textbf{Direction setting and corpus prep.} In line with STGT guidelines, we framed a single guiding question: \textbf{\emph{What are the lived experiences of women teaching support staff in IT/software engineering higher education, viewed through an EDI lens?}} This broad framing allowed for comprehensive exploration of the multifaceted experiences of women TSS in SE/IT education.
Given the first author's experience in technical education, working as a TSS in computing education, previous experience of research in the gender dynamics in STEM fields, we adopted a constructivist paradigm~\cite{charmaz2017power}. This paradigm acknowledges the researcher's active role in shaping the interview process, interpreting responses, and developing concepts based on participants' lived experiences.

\item \textbf{Open, incident-level coding (hashtag coding).} The first author conducted initial open coding on all transcripts using the hashtag coding technique in STGT, developing a preliminary codebook\footnote{While the raw data and codebook cannot be shared as per the ethics approval requirements, we provide examples of data analysis and multiple quotations throughout to ground the findings in empirical evidence.} \cite[Chapter 10]{hoda2024qualitative}. The first author coded line-by-line using STGT’s hashtag technique, keeping codes close to participants’ wording (e.g.,  \#AggressiveCommunicationByMaleStudentsTowardWomenTSS, \#MaleTSSHaveMorePerceivedAuthority). The unit of analysis was a concrete \emph{incident} (e.g., a mark dispute, classroom interruption, plagiarism confrontation). An illustrative coding snapshot is shown in Figure~\ref{fig:ExampleCoding}. For this example figure, we have focused on the codes related to student facing dynamics category.

\item \textbf{Clustering to concepts, subcategories, and categories (constant comparison).} Through constant comparison within and across transcripts, similar codes were grouped to form concepts, which were then organised into subcategories and categories (see the example in Figure~\ref{fig:ExampleCoding} for reference). The research team met regularly to review and refine the emerging conceptual categories. Visual mapping of relationships between concepts helped identify patterns and theoretical connections. Earlier interviews were revisited whenever codes or meanings shifted. Analytical memos were maintained throughout to capture researcher reflections and emerging insights. 

\item \textbf{Emergent intersectionality overlays.} To examine how gendered frictions intensified or attenuated with other identities, we applied analytic tags as contextual lenses (e.g., English as an additional language, perceived seniority/age in the classroom, prior industry experience, teaching tenure, local culture capital). We used these tags for co-occurrence checks and cross-case contrasts. See the example hashtag code in Figure~\ref{fig:ExampleCoding} highlighted in green, which refers to intersectionality. 

\item \textbf{Wrapping up and reporting conventions.} After 15 interviews, additional data seemed to elaborate existing categories rather than introduce new ones, and we assessed the corpus as sufficient for an experience-grounded account. 
\end{enumerate}

\subsubsection{Ensuring Rigour and Trustworthiness}
\textcolor{black}{The first author conducted all interviews and led the initial coding. This author has prior experience in SE/IT teaching support and EDI-focused research, which supported rapport-building and sensitivity to participants' accounts, but may also introduce interpretive bias (e.g., heightened importance of gendered dynamics). To mitigate this risk and ensure rigour and trustworthiness, we employed several strategies: (i) we maintained a detailed audit trail documenting all analytical decisions, including how codes were developed, merged, or refined. Regular team meetings ensured consistency: (ii) all authors reviewed the codes, concepts and categories to minimise individual bias. Disagreements were resolved through discussion and returning to the raw data; (iii) We maintained memos that provided rich descriptions of participants' lived experiences through extensive use of direct quotes. This systematic application of the STGT basic stage allowed us to develop a nuanced understanding of the multifaceted lived experiences of women TSS in SE/IT higher education.}

%% file: Tables/tab_participants.tex
\begin{table}[!htb]
\centering
\caption{Participant Experience, Level, and Specialisation}
\label{table:experience_level_specialisation}
\vspace*{-1em}
\footnotesize
\rowcolors{2}{BrandPink}{white} % start striping from row 2 (i.e., first data row)
\begin{tabular}{p{0.07\linewidth}p{0.13\linewidth}p{0.1\linewidth}p{0.5\linewidth}}
\hline
\textbf{Code} & \textbf{Teaching Exp.(Yrs)} & \textbf{UG/PG Level} & \textbf{Teaching Specialisation}\\
\hline
P1 & 2 & Both & Project Management, Software Dev. \\ \hline
P2 & 8 & UG & Project Management  \\ \hline
P3 & 2.5 & UG & IT/SE Capstone Projects \\ \hline
P4 & 2 & Both & Intro. to Programming\\ \hline
P5 & 3 & Both & Intro. to Programming, Systems Dev. \\ \hline
P6 & 4 & UG & Project Management, Student  Industry Experience, Business Information Systems \\ \hline
P7 & 15 & UG & Gaming \& Design \\ \hline
P8 & 5 & UG & Interactive Media (HCI) \\ \hline
P9 & 0.5 & Both & Intro. to Programming, Algorithms \\ \hline
P10 & 2 & Both & Software Engineering, Software Testing, Data Science \\ \hline
P11 & 1 & Both & Intro. to Programming, Cybersecurity \\
P12 & 7 & PG & Student Indutry Experience \\ \hline
P13 & 1.5 & UG & Project Management, Business Information Systems \\ \hline
P14 & 4.5 & UG & Gaming \& Design \\ \hline
P15 & 16 & PG & Software Industry Professional Practice \\
\hline
\end{tabular}
\vspace*{-1em}
\end{table}

%% file: Sections/discussion.tex
\section{Findings}~\label{sec:Findings}

In this section, we present insights from the experiences of the 15 women TSS, tracing how instructional authority for them is built, challenged, and maintained across teaching teams and in student-facing interactions with an EDI lens. We begin with \textbf{Voice among colleagues} and reflect on the (i) \emph{Losing Voice} experiences of public dismissal or interruption in classrooms and meetings, and back-channel override or withholding of support by colleagues; followed by (ii) \emph{Finding Voice} experiences, where explicit allyship, inclusive team routines, and mutual mentoring actively amplify women TSS contributions. 
Next, we turn to the classroom \textbf{Student-facing dynamics}, wherein we first discuss what participants describe as the ``payoff'' of teaching, i.e., rewarding connections built through approachability and empathetic scaffolding. We then examine \emph{Gendered Friction} in classrooms, i.e., dismissal and disrespect, mark disputes and authority challenges, classroom struggles, and academic-integrity standoff—moments where the same instruction or sanction carries less weight from a woman than when echoed by a man. Finally, we attend to \textbf{Intersectional positioning}, where we show that these patterns are not one-dimensional. Drawing on the wheel of power and privilege, we map how race/ethnicity, language capital (ESL/accent), teaching/industry experience, and perceived age interact with gender to intensify or reduce resistance. We use participant identifiers (P1–P15) as pseudonymous, and participants' quotations are reproduced verbatim with minor edits for clarity. We appreciate that the stories shared may be triggering for some. We have taken care to report participants' lived experiences as sensitively as possible without washing down their authenticity.

\subsection{Losing Voice: Negative Lived Experiences with Colleagues}
Across the 15 interviews, women TSS described a recurrent pattern in which male and occasionally more senior women colleagues subtly undermined their instructional authority. The incidents cluster into two overlapping subcategories: (i) \textit{public dismissal} in classrooms or team meetings, and (ii) \textit{back-channel override} of decisions or withholding of collegial help or support, with an underlying feeling of not being treated equal by colleagues.

\sectopic{Public dismissal and interruption.}
Several interviewees recalled being cut off mid-sentence or having their ideas waved aside during classes and team meetings. P1, recounting her first semester, said:

\TAquote{125}{While I was teaching, a male colleague jumped in and started pointing out what he thought I’d missed. In class, you’re supposed to finish your instructions and then invite your co-tutor to add anything. Interrupting someone mid-sentence breaks that basic etiquette -- and that’s exactly what [repeatedly] happened to me in my first semester. \ding{126}[P1]}

P7 experienced the same dynamic in a curriculum-design meeting: an idea rejected when voiced by her, accepted minutes later when a male tutor re-packaged it. She reflected (with much frustration),

\TAquote{125}{We were brainstorming ideas for a student assignment, and I suggested something. The [female] lecturer immediately dismissed it -- “No, that wouldn’t work.” But about 5 minutes later, a male TA suggested almost exactly the same thing, and suddenly it became a viable option [...] It was frustrating -- like the idea only became valid because it came from someone [with a male biological description]. \ding{126}[P7]}

The credibility of women TSS, as reported by the participants, is frequently questioned and undermined by their male colleagues. While talking about a temporary promotion to manage a course's teaching due to another lecturer's maternity leave, P13 recalled several instances of open dismissal by a male colleague, questioning, \textit{``How did you get this job?''}. 

\TAquote{125}{...some of my [male] colleagues were not pleased. They felt it should have been them [...] I could see that their unhappiness was growing. I noticed they undermined my authority by subtly telling students that [...] ``Your lecturer [P13] doesn’t know what she’s talking about. This assignment is bad. The feedback you’ve received is all wrong.'' \ding{126}[P13]}

\sectopic{Back-channel override or withholding support.}
A second mechanism through which peers undermined women TSS standing was back-channel override: private, unannounced changes to the women TSS academic decisions or simply withholding support to make the women staff members feel ``exposed''. P1 described marking a student's work only to discover that a male co-tutor had privately altered the feedback:

\TAquote{125}{Instead of consulting me [about the marks], he [the male co-tutor] gave the student additional feedback without asking. He implied that I was wrong, but rather than informing me about his conversation with the student, he just changed the feedback [in the assessment records] without my knowledge. \ding{126}[P1]}

P13 shared an experience while on medical leave, with the same male colleague (as the one who publicly dismissed her when she was temporarily promoted). 
\TAquote{125}{At the end of the teaching period, I had to go on medical leave for surgery. On the day I started my leave, this person sent an email to the entire team saying they were overloaded and couldn’t help with the marking that I was unable to fulfill. It felt like a betrayal, and I was disappointed because I didn’t expect him to let the team down.\ding{126}[P13]}

\begin{tcolorbox}[mybox] 
\textbf{Losing Voice: Negative Lived Experiences with Colleagues}. Participants' accounts show that women TSS \emph{lose voice} when colleagues interrupt, repackage, or quietly override their decisions—relegating them from co-educators to ``back-office'' task-doers and forcing additional emotional and administrative labour to defend routine judgments. Back-channel override and non-support combine to erode confidence, lengthen task completion times, and in cases push women TSS to refuse future teaching allocations with such colleagues.
\end{tcolorbox}

\subsection{Finding Voice: Positive Lived Experiences with Colleagues}
Against this backdrop of silencing, several participants also recalled moments, and sometimes whole teams in which, where their professional voice was \emph{actively amplified}.  Three peer practices underpin these positive experiences: explicit allyship that calls out bias in real time; inclusive micro-cultures deliberately fostered by leaders; and mutual mentoring that shares workload and confidence.

\TAquote{125}{[Redacted name of a male leader] was actually one of the first people to recognise my voice. I always tell him he's responsible for me being here because, before that, I never felt like I had a voice or my opinion was important. He was the one who decided I should take on a leadership role. It was small, just being a head tutor, but he encouraged me. If people were discussing an idea, he would ask for my opinion and didn't dismiss it. [...] It was the first time I felt heard. I would say that my entire career and what I can do today is thanks to him and the way he encouraged me to express myself.\ding{126}[P7]}

\sectopic{Allyship that calls out bias.}
On several occasions, a colleague—most often male—publicly called a student's/peer's behaviour as unacceptable, signalling that women's judgement carried equal weight. 

\TAquote{125}{He [a male colleague] was saying [to a disrespectful male student in a design class], ``You really shouldn’t be holding onto these outdated attitudes [...] Women are just as capable, and this [female] tutor has been working in the industry almost as long as I have. You should be giving her the same respect that you’re giving me.'' \ding{126}[P7]}

Such interventions halted the disrespect in the moment and for the rest of the semester towards the women TSS.

\sectopic{Inclusive team cultures set by strong leaders.}
Several participants worked in teaching teams where senior TSS or course leaders modelled praise, equal turn-taking, and gentle correction of errors.  P2 described the everyday rituals that normalised voice for all staff members, but particularly helpful for newcomers:

\TAquote{125}{In our teaching team, everyone is really respectful of each other. We don't talk over one another or try to make anyone look bad [in classroom], even if someone is inexperienced. In fact, we do the opposite. For example, we might say something like, ``Surprise! This is their first time teaching this tutorial, and they're already doing amazing.” We encourage the students to give them a round of applause and support. We try to make every TA feel appreciated for what they contribute. If someone says something that’s not entirely accurate or comprehensive, we just support them and gently clarify. [...] I think that’s been made possible because of the leadership. \ding{126}[P2]}

The above quote in contrast to quotes on losing voices, show how strong leaders build inclusive team cultures. Leaders also intervened when conflict loomed. In a large programming unit, P4 recounted:

\TAquote{125}{The head tutor, she’s also female, stepped in and said we needed to stop and redirect the student [who was being very argumentative over the marking of an assignment] to the head tutor and the lecturer [over their concerns]. Later she posted a message to all tutors and students about showing respect [towards all teaching staff]. \ding{126}[P4]}

These overt gestures legitimised women TSS authority and provided procedural cover for possible future confrontations.

\sectopic{Mutual mentoring and shared workload.}
Voices were further strengthened in teams that treated teaching as a shared craft rather than an individual test.  When illness struck, P1’s co-tutors redistributed marking without complaint:

\TAquote{125}{My team helped me finish the marking [...] I couldn’t attend in person [the teaching session being sick], and my chief examiner sent me a link to join online. Since I was sick and only a casual TA [contract type], I didn't want to miss anything. These people [the team and the senior colleagues] kept on motivating me and giving me positive vibes, which helps me feel like I can succeed. \ding{126}[P1]}

Day-to-day collaboration followed the same logic.  If a student ignored her explanation or ``encounter[ed] a difficult student or when the student isn't receptive to my advice'', P5 simply ``pulled in'' a co-tutor to re-phrase the point or handle a difficult student, framing collaboration as normal practice rather than a sign of weakness. Unlike the back-channel overrides described earlier, this was tutor-led and transparent: she invited a colleague in the moment, introduced them to the student, and stayed in the exchange, so the second voice amplified—rather than displaced—her judgement.

\begin{tcolorbox}[mybox] 
\textbf{Finding Voice: Positive Lived Experiences
with Colleagues.} Collectively, these experiences show that women TSS \emph{find voice} when peers and leaders:  
1) name bias aloud,  
2) embed respect in team routines, and  
3) treat teaching challenges as shared problems. Such practices offer concrete, peer-level levers to counter the negative mechanisms documented earlier.
\end{tcolorbox}

% ---------- Students ----------
\subsection{Rewarding Connections: Positive Lived Experiences With Students}

Across all interviews, the participants were unanimous on one point: their primary source of energy comes from watching students learn, gain confidence, and ``grow up'' academically and professionally.  ``Light bulb'' moments when a concept clicks, when a shy student finds their voice, or when learners connect classroom theory to future careers were repeatedly described as the most rewarding aspects of the role. P6 captured the sentiment articulately: 

\TAquote{125}{I think it's all about the connection with the students. After class, students approach me and say, ``Thank you so much! I really enjoyed this and all the help you've given me.'' I find that phenomenal. Which 20-year-old thinks to go up to their TA and express gratitude? [...] I really appreciate that connection. I've also had some great conversations about their lives outside of class. I enjoy being in the classroom and helping them make logical connections [in concepts]. \ding{126}[P6]}

Building those connections with students, however, does not happen automatically; it rests on deliberate rapport‐building, tailored personalised explanations, and the willingness to mentor beyond what the curriculum or norms dictate. The positive experiences below illustrate how TSS nurture that growth.

\sectopic{Approachability and empathetic support.}
Several staff described small, informal interactions, e.g., word games in class, quick check-ins at breaks—that open the door to deeper learning conversations.  
These micro-moments, they said, signal to students that questions are welcomed and that the teaching staff is ‘on their side’:

\TAquote{125}{One student this morning wanted to play a game with me: a small word game.  
These little interactions are actually things I enjoy. \ding{126}[P1]}

For some, the bond with students extends well past graduation, manifesting in heartfelt gestures that underline mutual respect:

\TAquote{125}{They give us a hug, make a cake for us with our photos on it, give us presents [...] that’s how much we have an influence on them. \ding{126}[P12]}

\sectopic{A gendered “empathy differential.”} Empathy also appears in how TSS adapt explanations.  
One interviewee routinely prepares extra ‘warm-up’ slides for those overwhelmed by first-year content, then walks the class through them step-by-step:

\TAquote{125}{When I put in a little extra effort, I receive positive feedback from students … they really appreciate it when we point out a few crucial slides and guide them to the next skill. \ding{126}[P11]}

Across interviews, women TSS consistently self-identified themselves as more approachable and patient than their male counterparts.  
Students would often queue for female TSS's time even when a male TSS was free in a classroom:

\TAquote{125}{I've observed that students often wait until we finish talking to another student, even when there are other male tutors available, just to ask us questions.  
It might be because they feel more comfortable approaching us. \ding{126}[P4]}

They also sought women out for ``deep-dive'' explanations:

\TAquote{125}{I think some students feel that female TAs are more willing to help, more patient, and likely to give detailed explanations.  
So they tend to ask me those detailed questions instead of asking male TAs.\ding{126}[P9]}

% Women tutors themselves noticed the contrast. \textcolor{red}{already said this above}
Staying back after class—or simply refusing to “rush off” -- became a quiet marker of care that some male colleagues did not match:

\TAquote{125}{I feel that, being women, I have more empathy than male tutors …  
For example, my [female] colleague and I would often stay back after class if students had doubts, but some of my male colleagues would say, “Oh, we have to rush,” and they’d just leave quickly.  
It’s those little instances of empathy that stand out. \ding{126}[P9]}

\begin{tcolorbox}[arc=0mm,width=\columnwidth,
                  top=1mm,left=1mm,  right=1mm, bottom=1mm,
                  boxrule=1pt, colback=green!5!white,colframe=green!75!black] 
~\textbf{Rewarding Connections: Positive Lived
Experiences With Students.} Taken together, these accounts suggest that \emph{approachability} and \emph{empathetic scaffolding} are not merely personal traits but are interpreted through a gendered lens: female TSS are both expected to and do invest more relational labour, i.e., the time and emotional effort to build rapport, monitor students’ affect, and tailoring explanations, which in turn shapes how students distribute their trust across the teaching team.
\end{tcolorbox}

% \faVenusMars

%--------------------------------------------------
\subsection{Gendered Friction: Negative Lived Experiences With Students}
Across the interviews, women TSS described a consistent pattern of \emph{gendered resistance}: male (and occasionally female) students were more likely to argue, disrupt, or ignore instructional authority when the TSS in front of them was a woman. Every participant could recall moments when this \emph{gendered resistance} from students surfaced. Across classrooms, labs, and inboxes, a consistent pattern emerged: the same statement, sanction, or request carried \emph{less weight} when delivered by a woman than when echoed by a man.

\sectopic{Dismissal and disrespect.}
Several women reported that pockets of (mostly male) students were quicker to discount their expertise or tune out their instructions, especially when no senior male was present to ``back them up.''  One participant described the latent assumption succinctly:
P7 traced the behaviour to lingering assumptions that ``women aren’t as smart as men'', which is potentially grounded in societal stereotypes that women are not as good in computing or technology-related fields~\cite{vitores2016trouble}.

\TAquote{125}{I do think some [male] students don’t feel that women are as worth listening to, or maybe they don’t give them the same level of respect as they do to male TAs. \ding{126}[P7]}

In its harsher form, the attitude tipped into outright rudeness, as in a workshop where a male student, challenged for late arrival by the woman TSS. As recounted by P15:

\TAquote{125}{One boy arrived half an hour, maybe 40 minutes late [to a student presentations' workshop]. He walked in without any apologies, just sat down. Then he got up to walk out, and I said, "Where are you going? You've just arrived." He looked at me and said, "Well, what do you want me to do? Piss myself? [suggesting they needed to use the restroom]" This is how they talk; this is the level of disrespect [mostly towards women TSS]. 
\ding{126}[P15]}

Such incidents happened infrequently but remained memorable, reinforcing women TSS sense that they started every class one step behind in earned credibility.

\sectopic{Mark disputes and authority challenges.}
Assessment was the most common trigger of conflict.  Women TSS described a ``two-step'' pattern: vigorous challenges followed by accusations of bias or even veiled threats when they delivered a contested mark, and then near-instant compliance when a male TSS later repeated the same judgement.
  
\TAquote{125}{A student complained to me for two weeks.  
The chief examiner (male) copied exactly what I had written; the student replied, “Okay, thanks,” and that was the end of it. \ding{126}[P7]}

\TAquote{125}{Male students tend to be a bit more argumentative. It seems that some of them perceive me as less authoritative compared to my male co-tutors. [...] I can feel that when they ask me questions, it often comes across as rude. I definitely feel they see me as less credible than my male co-tutors, which might explain why they feel more emboldened to challenge me about their marks. \ding{126}[P5]}

P11 received an email threatening escalation from a male student:
\TAquote{125}{
I confirmed that the mark I gave him was based on our marking criteria and was very objective. However, he was not happy with it. In his email, he mentioned that if this issue is not handled properly, he will ``escalate'' it. I feel this is a threat, and I am not happy about it. Therefore, I cc'd this to our chief examiner. [...]]
 This is the first time I had experienced such a serious issue. I also believe that if the student were communicating with my male colleagues he wouldn't have drafted the email in this tone. He probably thinks I'm just a small tutor, so he can threaten me, expecting a better mark. Otherwise, he will report me to my manager. 
 \ding{126}[P11]
}

\sectopic{Classroom control and disruptive behaviour.}
Vocal authority also mattered.  Tutors with softer voices—or simply perceived as softer—found it harder to silence side-conversations or keep students from drifting out mid-session:

\TAquote{125}{I basically have to bellow at them to get them to quiet down … I very rarely see this happen with male TAs, but yes, I do sometimes see it with female TAs. \ding{126}[P7]}

The differential showed up even in non-verbal cues: students walked out on a female TA mid-explanation yet stayed seated for the same activity led by a man:

\TAquote{125}{When \emph{I} was speaking, about 80\% of the class got up and left; later that week, most of them stayed and listened while a male TA delivered the very same content. \ding{126}[P8]}

\sectopic{Academic integrity standoffs.}
Finally, women TSS reported being treated as softer targets in plagiarism or cheating disputes.  
Students denied wrongdoing until a male TA intervened:

\TAquote{125}{Despite clear evidence, the student kept denying it.  
A male TA walked in, asked the same questions, and the student immediately apologised and admitted the work wasn’t theirs. \ding{126}[P8]} 

Patterns were similar for plagiarism cases more broadly:

\TAquote{125}{For three years in a row I’ve had \emph{more} plagiarism cases [attempts at cheating] than the two classes taught by male colleagues. They [the students] seem to think they can get away with it with me. \ding{126}[P14]}

\begin{tcolorbox}[mybox] 
~\textbf{Gendered Friction: Negative Lived
Experiences With Students.} These episodes reveal how student challenges cluster around moments where instructional \emph{authority} is most visible: holding the floor, assigning marks, and policing rules.  
Each point of friction is subtly gendered: women TSS are presumed friendlier, less experienced, or easier to push around, requiring them to invest additional emotional and disciplinary labour simply to achieve parity with male counterparts.
\end{tcolorbox}

\begin{figure}[t]
  \centering
  \includegraphics[width=0.95\linewidth]{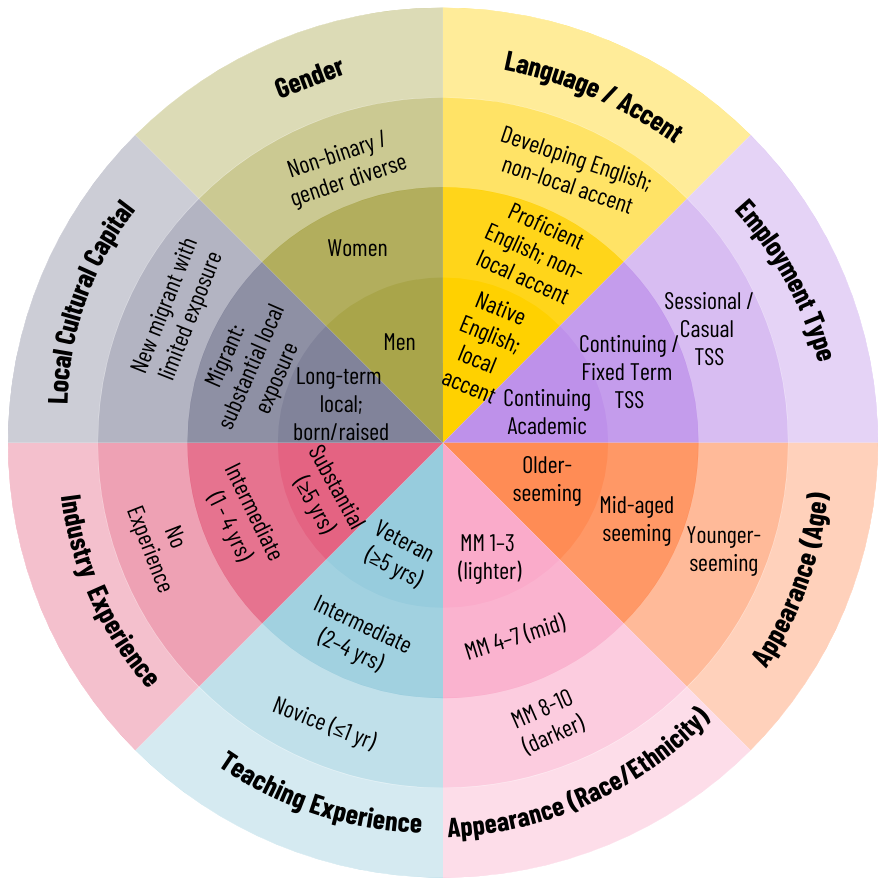}
  \vspace*{-1em}
  \caption[Intersectionality wheel of power and privilege in TSS teaching]%
  {Intersectionality wheel of privilege and power in TSS
   (adapted from the Canadian Council for Refugees).\protect\footnotemark}  \label{fig:wheel}
   \vspace*{-2em}
\end{figure}

\subsection{Intersectionality: Layered Identities} 
One of the key outcomes from our data analysis is that women TSS’ experiences with students, and colleagues (incl. leaders) are not governed by a single axis of gender.  Instead, \emph{who} attracts resistance and \emph{whose} expertise and authority is taken for granted shifts along multiple, intertwining dimensions—native‐language status, ``whiteness,'' teaching seniority, industry experience, perceived age based on appearance, accent, and more. This aligns with Kimberlé Crenshaw’s explanation that gender is only one coordinate on a matrix of ``overlapping or intersecting social identities'' that mediate access to power and exposure to bias~\cite{crenshaw2013mapping}. In Figure~\ref{fig:wheel}, we adapt the well-known \textit{wheel of power and privilege}\footnote{\url{https://www.canada.ca/content/dam/ircc/documents/pdf/english/corporate/anti-racism/wheel-privilege-power.pdf}} to visualise the most salient gradations encountered in our interviews, and use aligned terminology throughout. Skin tone of the participants is coded using the Massey–Martin 1–10 scale and, for reporting, collapsed to MM 1–3 (lighter), 4–7 (mid), and 8–10 (darker)~\cite{massey2003nis}.
Figure~\ref{fig:wheel} places a \emph{white (MM level 1-3) local, native‐English speaking, senior, male lecturer} at the centre (greatest automatic legitimacy) and radiates outward to combinations that our participants occupied. The farther a TSS identity markers sat from the centre, the more self-protective labour she reported doing. This figure and the analysis below is targeted to discuss that authority in the classroom is negotiated at the intersection of gender, language/accent, employment type, perceived age, race, teaching/industry experience, and the local culture capital. The wheel metaphor helps visualise why two women can hold the same role yet encounter vastly different levels of friction: every outward step from the centre adds another layer of credibility that needs to be earned. Below, we present an overview of some of these dimensions more explicitly covered in our interviews. Due to space limitations, we are unable to cover all the dimensions in detail.

%------------------------------------------------------

%------------------------------------------------------
\footnotetext{The original intersectionality wheel of privilege and power was developed by Sylvia Duckworth and Olena Hankivsky for the Canadian Council for Refugees.}

\sectopic{A. Race and ethnicity: “Not white?  Expect more push-back.”}
Several women noted that \emph{non-white} tutors, especially those who were also junior, received noticeably more remark requests and had their feedback second-guessed:

\TAquote{125}{
We get more re-mark requests for TAs who are not white. We have a few TAs from India or Sri Lanka, and they seem to get a lot more feedback challenges than the white TAs do. [...]
So yeah, I think it’s not just a gender issue—it’s probably also about race or ethnicity. Some students seem less likely to listen to someone who’s not white.
 \ding{126}[P7]}

\sectopic{B. Language capital:  Accent as a credibility test.}
An foreign accented TSS or one developing English proficiency deepened the scepticism some students already had about gender:

\TAquote{125}{Sometimes I feel that if I were a native English speaker, I could tell more jokes, be funnier, and make students feel more comfortable. That's something I feel is missing [...]
I feel that language is a barrier for me, and as I go along, I realise I need constant improvement in that area. \ding{126}[P10]}

\sectopic{C. Experience: Seniority speaks louder than gender, sometimes.}
Years in the teaching role and visible industry credentials could partly offset gendered doubt:

\TAquote{125}{I wouldn’t be surprised if there were two new TAs, one male and one female; students might lean toward the male.  
But my gender becomes less of an issue because of my experience. \ding{126}[P6]}

Yet women who lacked an experience level felt \emph{hyper-scrutinised}:

\TAquote{125}{In the first three weeks—female, young—there’s this attitude of “What do you know?”  Once we build rapport it’s fine, but early on there’s no [default] trust. \ding{126}[P12]}

\sectopic{D. Perceived age and appearance:  Youth = lesser authority.}
Younger-looking women were read as ``peers'' rather than instructors and paid a price in discipline encounters:

\TAquote{125}{In the programming unit, there were a couple of male students who would talk in a rude way. [...] I was a bit younger than them, their directness felt inappropriate for that setting. They would expect me to provide answers, and if I couldn't [...] they would respond in a somewhat rude manner. I felt that this was different from what I observed with the more mature or other male tutors.
\ding{126}[P4]}

P13’s lived experience below perfectly illustrates how authority is negotiated through intersecting status cues rather than gender alone. In her account, perceived youthfulness, a non-white appearance, and female gender combine to produce what participants elsewhere described as a credibility discount. Importantly, P13 attributes this to implicit, culturally learned deference rather than deliberate disrespect. This quote from P13 shows how in intersectional terms, the same woman TSS is read through multiple lenses at once (gender × ethnicity × perceived age)\footnote{We use the multiplier symbol instead of the addition symbol to denote the compounding effect of intersectionality.}, and those lenses compound to shape who is presumed competent and in charge.

\TAquote{125}{
I'm not that young; I'm 40 now, but I look younger than my actual age. [...] I'm from a Middle Eastern background. They [students] may have this unconscious bias towards age. If you are younger, it often means you should naturally be in a lower position in the hierarchy; that's an ingrained situation. The students I struggled with typically had that mindset. Being female added to that dynamic. [...] I don't think they do this intentionally [students being disrespectful], but when they see me in class alongside a male colleague, they often assume the male colleague is in charge. The older the male colleague, the more likely they are to unconsciously believe he is the authority figure. Some students call me ``Miss'' like they would a kindergarten teacher, even though I hold a PhD. I tend not to correct them, but it is quite disturbing to be called ``Miss'' while the male colleague, despite lacking a PhD, is referred to as ``Professor''.\ding{126}[P13]}

\begin{tcolorbox}[mybox] 
~\textbf{Intersectionality: Layered Identities.} Who gets called ``Miss'' and who gets heard isn't random: default legitimacy tracks an intersectional profile, such as, language, race, age cues, contract type and experience. Each step toward the outer ring multiplies the invisible labour required—more boundary-setting, more documentation, more backup—just to make the same instruction land with the weight it carries when someone closer to the centre says it.
While \emph{gender} is the main focus, advantages on other \emph{axes}—native English x local cultural capital x seniority x a permanent role x industry status—pull a TSS's \emph{radar profile toward the centre} (greater legitimacy); disadvantages push the profile \emph{outward}, increasing the relational, emotional, and disciplinary labour needed to reach parity. 
\end{tcolorbox}

\sectopic{How intersectionality works.}
In our data, \emph{authority is negotiated at the intersections} of language capital, local cultural familiarity, employment status, perceived age, race/ethnicity, and experience—so the same woman can occupy different positions on the “wheel” from class to class. Most participants sat on \emph{multiple outer axes} simultaneously (e.g., female x Asian x non-native english speaker x sessional x younger‐seeming), and reported doing extra, often invisible, work to offset credibility gaps:

\begin{itemize}
\item writing \textbf{longer, more anticipatory feedback} to pre-empt challenges and mark disputes,
\item \textbf{escalating or co-teaching with a senior/male ally} when push-back hardened or turned hostile,
\item investing additional \textbf{relationship-building} (check-ins, tone-setting, clear boundaries) to earn the same baseline respect.
\end{itemize}

\begin{figure}[t]
  \centering
  \includegraphics[width=0.9\linewidth]{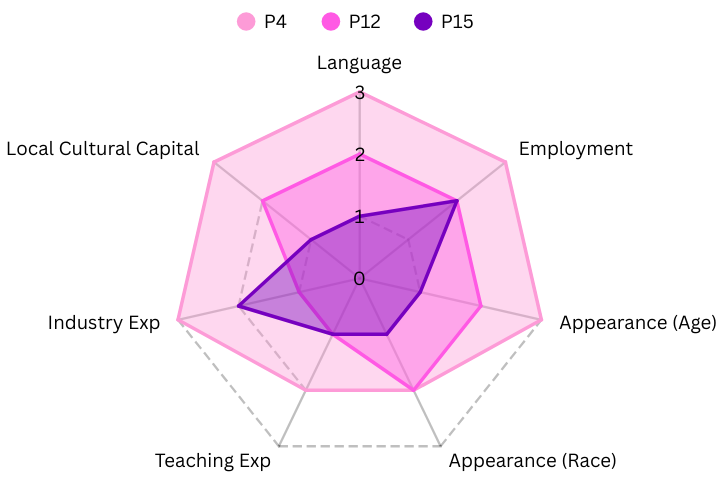}
  \caption%
  {Example Privilege and Power Radar}
  \label{fig:exampleRadar}
\begin{minipage}{0.5\textwidth}\footnotesize
\emph{Legend (dimensions).}  
\textbf{Language}: 1=native English; 2=fluent English (non-local accent); 3=developing English (foreign accent). \\
\textbf{Employment}: 1=tenured academic; 2=tenured TSS; 3=sessional TSS. \\
\textbf{Appearance (Age)}: 1=older seeming; 2=mid-aged seeming; 3=younger seeming. \\
\textbf{Appearace (Race)} : 1=White; 2=Different Shades; 3=Dark. \\
\textbf{Teaching Exp}: 1=5+ years; 2=2-4 years; 3=1 or less years. \\
\textbf{Industry Exp}: 1=5+ years; 2=1-4 years; 3 = 0 years. \\
\textbf{Local Cultural Capital} (\countryName): 1=Born / Raised; 2=Migrant with close local cultural understanding; 3 = Migrant with little or not close cultural understanding.
\end{minipage}
\vspace*{-1.5em}
\end{figure}

Figure~\ref{fig:exampleRadar} illustrates three example participants' cases - P4, P12 and P15. P4 sits furthest from the centre (sessional x younger-seeming x developing local/linguistic capital x limited industry experience) and thus absorbs the heaviest relational and disciplinary labour: boundary-testing by (mostly male) students, argumentative mark appeals, and approachability misread as pliability that requires constant redirection to seniors. P12 appears mid-ring but is \emph{re-centred} by high-status in extensive industry/teaching experience and a permanent, education-focused role.
P15 is identity-privileged (white x native English x older x long-serving x locally acculturated), yet large, noisy rooms and unit politics can still pull her outward: she reported confrontation against open defiance during presentations and sometimes seeks safety support, and her voice is muted unless leadership is explicitly inclusive.

%% file: Sections/recommendations.tex
\section{Recommendations to Strengthen Women TSS Authority and Experience}~\label{sec:Recommendations}
% : Multi–Level Actions

Below we discuss high and multi–level recommendations based on the patterns across interviews, our intersectionality analysis (incl. memos) and the existing literature. Overall, in addition to gender, legitimacy in classrooms is negotiated across language, local cultural capital, employment security, perceived age, race/ethnicity, and teaching/industry experience. Policies and routines should be \emph{intersectionally designed}, embedding authority structurally rather than women continuously re-earning it.

\subsection*{A. For University \& Faculty Leadership}
\begin{enumerate}[label=A\arabic*.,leftmargin=*]
  \item \textbf{Set norms and back them with policy.} Ambiguity about acceptable conduct leaves women TSS, espl. those in outer–ring positions (e.g., accented, younger–seeming, sessional) to police boundaries of acceptable behaviour alone. Publish and enforce a student–staff conduct code~\cite{TEQSA_Wellbeing} covering classes, forums, and email that explicitly names gendered disrespect, microaggressions, and mark–dispute harassment as breaches, with clear reporting channels and consequences.
  \item \textbf{Resource safety and de-escalation.} TSS regularly face heated disputes without training, with risks amplified for outer–ring staff. Fund recurring training in de-escalation~\cite{mirza2019undergraduate}, boundary-setting scripts, and academic-integrity procedures; guarantee after-hours security and a clear pathway for embedding safety.
  \item \textbf{Recognise relational and emotional labour.} Mentoring, mediation, and incident documentation remain invisible in workload models~\cite{crisp2022academic}, disproportionately burdening women TSS. Add explicit allowances for relational and emotional labour and such that paid marking time reflects actual effort.
  \item \textbf{Stabilise sessional work.} Sessional teaching contracts push outer-ring TSS furthest from power, weakening authority and retention. Establish transparent pathways from sessional to fixed-term/continuing (e.g., education-focused tracks)~\cite{smithers2025decasualisation}, and prioritise their access to professional development and consistent allocations. These help improve confidence in women TSS, as told by P14:
  \textit{``I think in the past, I was a bit worried about calling a student for bad behavior. Now, I feel more confident [about my job security after getting a tenured role].''}
\end{enumerate}

\subsection*{B. For Program/Course Leadership}
\begin{enumerate}[label=B\arabic*.,leftmargin=*]
  \item \textbf{Design for allyship.} Credibility gaps widen when staffing ignores intersectional profiles.  Pair tutors strategically (e.g., sessional/younger–seeming with permanent x experienced x local) in teaching sessions, and rotate visible leadership.
  \item \textbf{Codify co-teaching etiquette.} Ad-hoc practices enable mid-sentence overrides and credit theft that silence women TSS, especially the outer ring women TSS. Adopt a written co-teaching protocol: scripted handovers, no mid-sentence overrides, public attribution of ideas, and a standing post-class debrief to correct any slips not caught in the moment.
  \item \textbf{Establish competence handshake.} A key issue for outer ring women TSS is students' perception of their expertise. An option could be to adopt a Week-1 ``competence handshake'': a brief, repeatable routine where TSS publicly demonstrate disciplinary expertise with explicit endorsement from course leadership, and rotate the lead role to TSS most likely to face credibility discounts, e.g, outer-ring women.
\end{enumerate}

\subsection*{C. For Teaching Teams and TSS}
\begin{enumerate}[label=C\arabic*.,leftmargin=*]
  \item \textbf{Protect your boundaries.} Persistent grade disputes and frequent integrity standoffs become personalised and disproportionately drain the emotional labour of (especially junior) women TSS. Maintain shared scripts for common pushbacks, set a clear handover point in the escalation ladder (tutor\,$\rightarrow$\,head tutor\,$\rightarrow$\,course leader), and always cc-in a colleague or leader on contentious threads to avoid one-to-one loops.

  \item \textbf{Document and escalate early.} Micro-aggressions and recurring disrespect often go unrecorded, making patterns hard to prove and TSS easily feeling isolated. Keep a brief incident log (date, context, outcome) in a shared tracker and escalate at the \emph{second} recurrence rather than absorbing repeat behaviour.
  \item \textbf{Leverage intersectional strengths.} Accents, industry experience, and cultural fluencies can be misread as deficits. Reframe these attributes as assets in examples and mentoring, and pair tutors strategically to address age/authority/culture gaps.
\end{enumerate}

\subsection*{D. For Students (as Partners)}
\begin{enumerate}[label=D\arabic*.,leftmargin=*]
\item \textbf{Practice professionalism.} Informality can slip into disrespect, especially toward outer–ring TSS (e.g., younger–seeming). Treat classrooms as professional spaces: use role‐appropriate forms of address (avoid gendered honorifics like ``Miss''), keep a civil tone, and ground disagreements in evidence and rubrics.
\item \textbf{Use fair appeal channels.} Personal lobbying and repeated emails to individual TSS amplify power imbalances. Submit grade concerns via the published re-mark process with specific rubric references and examples, and direct correspondence to the course inbox/CE rather than individual staff.
\item \textbf{Be an ally in the room.} Unchecked microaggressions (interruptions, tone) silence TSS and peers. If you witness it, model good norms (``Let’s hear them finish''), support the posted etiquette, and respect when a TSS sets a boundary.
\end{enumerate}

%% file: Sections/background.tex
\section{Related Work}\label{sec:Background}

Related works have examined EDI in IT/SE education, with a sustained emphasis on gendered participation and climate \cite{margolis2002unlocking,rodriguez2021perceived,browning2024queering,hyrynsalmi2025challenges,de2025diversity,toti2025diversity}. Most references below cover issues related to women, however, some studies, e.g.,~\cite{fitch2003not}, do not specifically focus on women.  By contrast, there are comparatively few studies that focus on the everyday conditions of TSS often discussed under the broader label of teaching assistants (TAs) in IT/SE programs, though adjacent work in general computing education highlights their pedagogical centrality and challenges~\cite{mirza2019undergraduate,riese2020teaching,patel2025exploration,riese2021challenges}. Across large STEM classes, TAs measurably influence learning, attitudes, and persistence beyond content delivery~\cite{alzen2018logistic,de2025teaching}; in CS specifically, they frequently operate as community facilitators who scaffold belonging and provide humanistic care alongside technical help~\cite{perlmutter2023field,barkhuff2025exploring}. \textit{Visibility} also matters: women TAs can function as proximal role models that counter stereotype threat and shift participation norms in male-dominated classrooms~\cite{griffith2021role}, and \textit{identity proximity} has been shown to bolster Asian women’s engagement and confidence in computing~\cite{tari2018someone,tari2021asian}.
At the same time, literature documents how bias erodes women TSS’ perceived legitimacy. Work on the International TAs (ITAs) focuses on accent, appearance, and perceived ``foreignness,'' critiquing classroom management and individualised attention~\cite{fitch2003not}. More recent studies trace gendered racism whereby non-white and accented women are marked as less ``professional,'' facing microaggressions and credibility discounts in daily interactions~\cite{ramjattan2023engineered,huang2023marginalized,hebbani2014capturing,jackson2025intersecting}. These judgments surface in formal metrics as well: female graduate TSS tend to receive lower student evaluations than male peers despite comparable performance, with penalties amplified where ``technical authority'' is stereotyped as masculine~\cite{khazan2019examining,evans2024gender}.

This lived experiences report adds three major contributions when positioned against the related work. First, the paper centers \emph{women TSS} embedded in SE/IT programs and foreground the specific micro-practices where authority is negotiated (mark disputes, classroom control, integrity cases), which prior work largely treats indirectly. Second, we attempt to operationalise intersectionality, i.e., language/accent, local cultural capital, employment security, perceived age, race/ethnicity, and teaching/industry experience, into a practical, multi-dimensional intersectional profile as wheel of power and privilege that explains \emph{when} and \emph{why} the same woman is granted different legitimacy across contexts, and how different women at different positions on the wheel are likely to experience the same educational contexts differently. Third, we translate patterns into actionable, multi-level recommendations.

%% file: Sections/threats.tex
\section{Limitations and Threats to Validity}~\label{sec:threats}
We discuss limitations and threats to validity of our study in this section.
Key constructs (e.g., ``gendered resistance,'' ``relational labour,'' and our intersectionality profile) are operationalised from interviews and prior frameworks (through targeted literature review after data collection and analysis). Nonetheless, they can simplify complex social processes. Our study relies on self–reports of women TSS, grounding in concrete examples of lived experiences (as opposed to opinions). {Also, this focus on women TSS alone limits triangulation; future work can incorporate students’ perspectives, observations, and accounts from other staff (e.g., male TSS) to corroborate and contextualise these dynamics.}

The radar visualisation is meant to be descriptive (not causal) and may misclassify edge cases (e.g., accent vs. language proficiency), so related findings should be read as patterns, not precise quantitative measurements. Causal attributions (e.g., that gender or accent caused specific student behaviours) cannot be established from interview data alone; confounding factors such as class/course size, assessment timing, cohort composition, and classroom layout may contribute. We focus in this paper on lived experiences instead of causal inference, and present verbatim quotes that show how inferences were drawn. Qualitative coding and interpretation are inherently researcher–mediated (using a constructivist approach); different analysts might cluster concepts differently. {The first author conducted the interviews and led initial coding. We mitigate this through reviews by multiple researchers and a consistent coding system (hashtags → concepts → subcategories → categories) and by linking each grouping to exemplar quotations.}
As an exploratory, qualitative, descriptive study with a sample of 15 women TSS in IT/SE at one university context, we do not claim generalisation. {Transferability may also vary because TSS role structures differ across educational systems and countries.} However, lived experienced studies provide valuable insights into the experiences of a specific cohort that those outside it may be oblivious to, through hard-earned data~\cite{seaman1999qualitative}. Additionally, our accounts echo patterns reported in adjacent computing/engineering contexts.

%% file: Sections/conclusion.tex
\section{Conclusion}~\label{sec:Conclusion}

This lived experiences report traced how instructional authority for women teaching–support staff (TSS) in information technology/software engineering higher education is built, challenged, and maintained. Across interviews, participants located the payoff of teaching in students’ growth—``light-bulb'' moments and durable mentoring relationships—sustained
through approachability and empathetic scaffolding. Yet the same accessibility often incurs a relational and emotional labour: compared with men, women TSS more frequently faced dismissal or interruption by colleagues, and in student-facing settings encountered gendered friction around dismissal/disrespect, mark disputes, classroom control, and academic integrity. We found that explicit allyship, inclusive micro–cultures, and mutual mentoring help women find voice, while back-channel overrides and public cutting-off erode it. Importantly, we demonstrated that these dynamics are not one-dimensional: our intersectionality analysis explains why the same woman is granted different default legitimacy across contexts and why different women can have varied experiences in the same contexts. We translate these
insights into multi-level recommendations that render authority structural rather than something women must continually re-earn.

In the future, we plan to run comparative studies with students, men and women TSS to map how each group defines similar issues highlighted in this paper. To account for student background in such studies, we plan to pair surveys and classroom observations with administrative data to model how country-of-origin socialisation and prior schooling influence acceptance of authority from women vs. men. Finally, we plant to apply our recommendations across multiple courses within our own teaching contexts to improve the lived experiences of women TSS, one of the most vulnerable roles in IT/SE higher education.

%% file: sample-base.bib
@article{seaman1999qualitative,
  title={Qualitative methods in empirical studies of software engineering},
  author={Seaman, Carolyn B.},
  journal={IEEE Transactions on software engineering},
  volume={25},
  number={4},
  pages={557--572},
  year={1999},
  publisher={IEEE}
}

@article{massey2003nis,
  title={The NIS skin color scale},
  author={Massey, Douglas S and Martin, Jennifer A},
  journal={Office of Population Research, Princeton University},
  year={2003}
}

@article{griffith2021role,
  title={The role of the teaching assistant: Female role models in the classroom},
  author={Griffith, Amanda L and Main, Joyce B},
  journal={Economics of Education Review},
  volume={85},
  pages={102179},
  year={2021},
  publisher={Elsevier}
}

@article{evans2024gender,
  title={Gender patterns in engineering PhD teaching assistant evaluations corroborate role congruity theory},
  author={Evans, CA and Adler, K and Yucalan, D and Schneider-Bentley, LM},
  journal={International Journal of STEM Education},
  volume={11},
  number={1},
  pages={5},
  year={2024},
  publisher={Springer}
}

@article{khazan2019examining,
  title={Examining gender bias in student evaluations of teaching for graduate teaching assistants},
  author={Khazan, E and Borden, J and Johnson, S and Greenhaw, L},
  journal={NACTA Journal},
  volume={64},
  pages={422--427},
  year={2019},
  publisher={JSTOR}
}

@inproceedings{jackson2025intersecting,
  title={Intersecting Identities: Graduate Teaching Assistant Experiences in the Academy},
  author={Jackson, Justina Rodriguez},
  booktitle={The Educational Forum},
  volume={89},
  number={2},
  pages={194--209},
  year={2025},
  organization={Taylor \& Francis}
}

@article{hebbani2014capturing,
  title={Capturing the experiences of international teaching assistants in the US American classroom},
  author={Hebbani, Aparna and Hendrix, Katherine Grace},
  journal={New Directions for Teaching and Learning},
  volume={2014},
  number={138},
  pages={61--72},
  year={2014},
  publisher={Wiley Online Library}
}

@article{fitch2003not,
  title={“Not a lick of English”: Constructing the ITA identity through student narratives},
  author={Fitch, Fred and Morgan, Susan E},
  journal={Communication Education},
  volume={52},
  number={3-4},
  pages={297--310},
  year={2003},
  publisher={Taylor \& Francis}
}

@inproceedings{tari2021asian,
  title={How Asian women’s intersecting identities impact experiences in introductory computing courses},
  author={Tari, Mina and Hua, Vivian and Ng, Lauren and Annabi, Hala},
  booktitle={International Conference on Information},
  pages={603--617},
  year={2021},
  organization={Springer}
}

@article{tari2018someone,
  title={Someone On My Level: How Women of Color Describe the Role of Teaching Assistants in Creating Inclusive Technology Courses},
  author={Tari, Mina and Annabi, Hala},
  year={2018}
}

@inproceedings{perlmutter2023field,
  title={" A field where you will be accepted": Belonging in student and TA interactions in post-secondary CS education},
  author={Perlmutter, Leah and Salac, Jean and Ko, Amy J},
  booktitle={Proceedings of the 2023 ACM Conference on International Computing Education Research-Volume 1},
  pages={356--370},
  year={2023}
}

@inproceedings{barkhuff2025exploring,
  title={Exploring the Humanistic Role of Computer Science Teaching Assistants across Diverse Institutions},
  author={Barkhuff, Grace and Pruitt, Ian and Namani, Vyshnavi and Johnson, William Gregory and Borela, Rodrigo and Zegura, Ellen and Bourgeois, Anu G and Shapiro, Ben Rydal},
  booktitle={Proceedings of the 56th ACM Technical Symposium on Computer Science Education V. 1},
  pages={67--73},
  year={2025}
}

@article{alzen2018logistic,
  title={A logistic regression investigation of the relationship between the Learning Assistant model and failure rates in introductory STEM courses},
  author={Alzen, Jessica L and Langdon, Laurie S and Otero, Valerie K},
  journal={International journal of STEM education},
  volume={5},
  number={1},
  pages={1--12},
  year={2018},
  publisher={Springer}
}

@inproceedings{patel2025exploration,
  title={Exploration of Undergraduate Teaching Assistant Identity and Teaching Goals in Data Science Courses},
  author={Patel, Krina and Brooks-Ramirez, Abigail and Dang, Rebecca and Adolfo Ventura Benitez, Bryan and Yan, Lisa},
  booktitle={Proceedings of the 56th ACM Technical Symposium on Computer Science Education V. 2},
  pages={1573--1574},
  year={2025}
}

@inproceedings{riese2020teaching,
  title={Teaching assistants’ experiences of tutoring and assessing in computer science education},
  author={Riese, Emma and Kann, Viggo},
  booktitle={2020 IEEE Frontiers in Education Conference (FIE)},
  pages={1--9},
  year={2020},
  organization={IEEE}
}

@inproceedings{browning2024queering,
  title={Queering software engineering education: integrative approaches and student experiences},
  author={Browning, Jonathan W and Bustard, John and Anderson, Neil},
  booktitle={2024 21st International Conference on Information Technology Based Higher Education and Training (ITHET)},
  pages={1--8},
  year={2024},
  organization={IEEE}
}

@incollection{toti2025diversity,
  title={Diversity, Equity, and Inclusion in Computing Science: Culture is the Key, Curriculum Contributes},
  author={Toti, Giulia and Lindner, Peggy and Gao, Alice and Baghban Karimi, Ouldooz and Engineer, Rutwa and Hur, Jinyoung and McNeill, Fiona and Reckinger, Shanon and Robinson, Rebecca and Sollazzo, Anna and others},
  booktitle={2024 Working Group Reports on Innovation and Technology in Computer Science Education},
  pages={175--225},
  year={2025}
}

@inproceedings{de2025diversity,
  title={Diversity in software engineering education: Exploring motivations, influences, and role models among undergraduate students},
  author={de Souza Santos, Ronnie and Santos, Italo and Santos, Robson and Magalhaes, Cleyton},
  booktitle={IEEE Conference on Software Engineering Education and Training (CSEE\&T 2025)},
  pages={1--12},
  year={2025}
}

@article{rodriguez2021perceived,
  title={Perceived diversity in software engineering: a systematic literature review},
  author={Rodr{\'\i}guez-P{\'e}rez, Gema and Nadri, Reza and Nagappan, Meiyappan},
  journal={Empirical Software Engineering},
  volume={26},
  number={5},
  pages={102},
  year={2021},
  publisher={Springer}
}

@article{hyrynsalmi2025challenges,
  title={Challenges and opportunities: Implementing diversity and inclusion in software engineering university level education in Finland},
  author={Hyrynsalmi, Sonja M},
  journal={Journal of Systems and Software},
  volume={219},
  pages={112239},
  year={2025},
  publisher={Elsevier}
}

@article{crisp2022academic,
  title={Academic workloads: what does a manager need to consider?},
  author={Crisp, Beth R},
  journal={Journal of Higher Education Policy and Management},
  volume={44},
  number={6},
  pages={547--562},
  year={2022},
  publisher={Taylor \& Francis}
}

@article{gunatilake2025role,
  title={The Role of Empathy in Software Engineering--A Socio-Technical Grounded Theory},
  author={Gunatilake, Hashini and Grundy, John and Hoda, Rashina and Mueller, Ingo},
  journal={ACM Transactions on Software Engineering (TOSEM)},
  year={2025}
}

@inproceedings{gama2025socio,
  title={A Socio-Technical Grounded Theory on the Effect of Cognitive Dysfunctions in the Performance of Software Developers with ADHD and Autism},
  author={Gama, Kiev and Liebel, Grischa and Goul{\~a}o, Miguel and Lacerda, Aline and Lacerda, Cristiana},
  booktitle={2025 IEEE/ACM 47th International Conference on Software Engineering: Software Engineering in Society (ICSE-SEIS)},
  pages={1--12},
  year={2025},
  organization={IEEE}
}

@article{de2025teaching,
  title={Teaching assistants’ contributions to creating inclusive and equitable learning spaces in engineering},
  author={de Lima, Joelyn and Isaac, Siara and Kovacs, Helena},
  journal={European Journal of Engineering Education},
  volume={50},
  number={1},
  pages={99--116},
  year={2025},
  publisher={Taylor \& Francis}
}

@article{smithers2025decasualisation,
  title={Decasualisation and the universities accord: an examination of university approaches},
  author={Smithers, Kathleen and Harris, Jess and Heffernan, Troy and Gurr, Sarah},
  journal={Journal of Higher Education Policy and Management},
  volume={47},
  number={3},
  pages={282--298},
  year={2025},
  publisher={Taylor \& Francis}
}

@misc{TEQSA_Wellbeing,
  author       = "TEQSA: Australian Tertiary Education Quality and Standards Agency",
  title        = "Guidance note: Wellbeing and safety",
  howpublished = "https://www.teqsa.gov.au/guides-resources/resources/guidance-notes/guidance-note-wellbeing-and-safety",
  year         = "2018"
}

@article{huang2023marginalized,
  title={Marginalized, silenced, and struggling: Understanding the plights of Chinese graduate teaching assistants},
  author={Huang, Ting and Chen, Si and Lin, Jia and Cun, Aijuan},
  journal={International Journal of Chinese Education},
  volume={12},
  number={1},
  pages={2212585X231156996},
  year={2023},
  publisher={SAGE Publications Sage UK: London, England}
}

@article{ramjattan2023engineered,
  title={Engineered accents: International teaching assistants and their microaggression learning in engineering departments},
  author={Ramjattan, Vijay A},
  journal={Teaching in Higher Education},
  volume={28},
  number={6},
  pages={1119--1134},
  year={2023},
  publisher={Taylor \& Francis}
}

@inproceedings{tari2019understanding,
  title={Understanding Undergraduate Teaching Assistants’ Perspectives on Inclusive Pedagogy in Introductory Computing Courses},
  author={Tari, Mina and Annabi, Hala and Ko, Andrew},
  booktitle={2019 Research on Equity and Sustained Participation in Engineering, Computing, and Technology (RESPECT)},
  pages={1--2},
  year={2019},
  organization={IEEE}
}

@incollection{crenshaw2013mapping,
  title={Mapping the margins: Intersectionality, identity politics, and violence against women of color},
  author={Crenshaw, Kimberl{\'e} Williams},
  booktitle={The public nature of private violence},
  pages={93--118},
  year={2013},
  publisher={Routledge}
}

@article{vitores2016trouble,
  title={The trouble with ‘women in computing’: a critical examination of the deployment of research on the gender gap in computer science},
  author={Vitores, Anna and Gil-Ju{\'a}rez, Adriana},
  journal={Journal of Gender Studies},
  volume={25},
  number={6},
  pages={666--680},
  year={2016},
  publisher={Taylor \& Francis}
}

@article{charmaz2017power,
  title={The power of constructivist grounded theory for critical inquiry},
  author={Charmaz, Kathy},
  journal={Qualitative inquiry},
  volume={23},
  number={1},
  pages={34--45},
  year={2017},
  publisher={SAGE Publications Sage CA: Los Angeles, CA}
}

@inproceedings{riese2021challenges,
  title={Challenges faced by teaching assistants in computer science education across europe},
  author={Riese, Emma and Lor{\aa}s, Madeleine and Ukrop, Martin and Effenberger, Tom{\'a}{\v{s}}},
  booktitle={Proceedings of the 26th ACM Conference on Innovation and Technology in Computer Science Education V. 1},
  pages={547--553},
  year={2021}
}

@inproceedings{mirza2019undergraduate,
  title={Undergraduate teaching assistants in computer science: a systematic literature review},
  author={Mirza, Diba and Conrad, Phillip T and Lloyd, Christian and Matni, Ziad and Gatin, Arthur},
  booktitle={Proceedings of the 2019 ACM Conference on International Computing Education Research},
  pages={31--40},
  year={2019}
}

@article{murciano2022missing,
  title={Missing women in tech: The labor market for highly skilled software engineers},
  author={Murciano-Goroff, Raviv},
  journal={Management Science},
  volume={68},
  number={5},
  pages={3262--3281},
  year={2022},
  publisher={INFORMS}
}

@article{george2024bridging,
  title={Bridging the gender gap in STEM: Empowering women as drivers of technological innovation},
  author={George, A Shaji},
  journal={Partners Universal Innovative Research Publication},
  volume={2},
  number={2},
  pages={89--105},
  year={2024}
}

@article{hoda2021socio,
  title={Socio-technical grounded theory for software engineering},
  author={Hoda, Rashina},
  journal={IEEE Transactions on Software Engineering},
  volume={48},
  number={10},
  pages={3808--3832},
  year={2021},
  publisher={IEEE}
}

@article{hoda2024qualitative,
  title={Qualitative research with socio-technical grounded theory},
  author={Hoda, Rashina},
  journal={Springer},
  year={2024},
  publisher={Springer}
}

@book{margolis2002unlocking,
  title={Unlocking the clubhouse: Women in computing},
  author={Margolis, Jane and Fisher, Allan},
  year={2002},
  publisher={MIT press}
}

@misc{google2016diversity,
  title={Diversity gaps in computer science: Exploring the underrepresentation of girls, blacks and hispanics},
  author={Google Inc. \& Gallup Inc.},
  year={2016},
  publisher={Google Inc. \& Gallup Inc.}
}

@article{sax2017anatomy,
  title={Anatomy of an enduring gender gap: The evolution of women’s participation in computer science},
  author={Sax, Linda J and Lehman, Kathleen J and Jacobs, Jerry A and Kanny, M Allison and Lim, Gloria and Monje-Paulson, Laura and Zimmerman, Hilary B},
  journal={The Journal of Higher Education},
  volume={88},
  number={2},
  pages={258--293},
  year={2017},
  publisher={Taylor \& Francis}
}

@inproceedings{wang2015gender,
  title={Gender differences in factors influencing pursuit of computer science and related fields},
  author={Wang, Jennifer and Hong, Hai and Ravitz, Jason and Ivory, Marielena},
  booktitle={Proceedings of the 2015 ACM Conference on Innovation and Technology in Computer Science Education},
  pages={117--122},
  year={2015}
}
